\documentclass[%
 reprint,
superscriptaddress,
 amsmath,amssymb,aps, prl, floatfix, showpacs
]{revtex4-1}

\usepackage{graphicx}
\usepackage{dcolumn}
\usepackage{bm}
\usepackage{amsfonts,amssymb,amsmath}
\usepackage[colorlinks,linkcolor={blue},citecolor={blue},urlcolor={blue}]{hyperref}
\usepackage{siunitx}
\usepackage{textcomp}
\usepackage{float}
\def\doubleunderline#1{\underline{\underline{#1}}}
\usepackage{color}
\begin{document}


\title{Mass inversion in graphene by proximity to dichalcogenide monolayer}

\author{Abdulrhman M. Alsharari}
 \email{aalsharari@ut.edu.sa}
 \affiliation{Department of Physics and Astronomy, and Nanoscale and Quantum
 	Phenomena Institute, \\ Ohio University, Athens, Ohio 45701}
 \affiliation{Department of Physics, University of Tabuk, Tabuk, 71491, SA}
\author{Mahmoud M. Asmar}
\affiliation{Department of Physics and Astronomy, Louisiana State University, Baton Rouge, LA 70803}
\author{Sergio E. Ulloa}
\affiliation{Department of Physics and Astronomy, and Nanoscale and Quantum
	Phenomena Institute, \\ Ohio University, Athens, Ohio 45701}
%

\date{\today}

\begin{abstract}
Proximity effects resulting from depositing a graphene layer on a TMD substrate layer change the dynamics of the electronic states in graphene,
inducing spin orbit coupling (SOC) and staggered potential effects.  An effective Hamiltonian
that describes different symmetry breaking terms in graphene, while preserving time reversal invariance, shows that an inverted mass band gap regime
is possible.  The competition of different perturbation terms causes a transition from an
inverted mass phase to a staggered gap in the bilayer heterostructure, as seen in its phase diagram.
A tight-binding calculation of the bilayer validates the effective model parameters.
A relative gate voltage between the layers may produce such phase transition in experimentally accessible
systems.  The phases are characterized in terms of Berry curvature and valley Chern numbers, demonstrating that the system
may exhibit quantum spin Hall and valley Hall effects.
\end{abstract} 

\maketitle

%
Graphene has many interesting properties intensively studied  in recent years \cite{graphene}. Prominent among these, possible intrinsic spin orbit
coupling (SOC) on its charge carriers was estimated by Kane and Mele to be rather weak, $\simeq 1$ \si\micro \si{eV} \cite{qshe}.
Improved estimates that include contributions from $d$-orbitals yield larger values, $\simeq$ 24 \si\micro \si{eV} \cite{gdft}, although still rather weak for experimental observation.
Several ways have been proposed to enhance the spin orbit interaction in graphene for uses in spintronics \cite{spintronics}.
Enhancing $ sp^3 $ hybridization by adding hydrogen or fluorene atoms \cite{fluor}, as well as decorating with heavy adatoms \cite{WU}, or
different substrates \cite{ind-g},
have been proposed to produce large SOC.\@
Depositing graphene on metallic substrates has also resulted in strong SOC for the charge carriers in graphene \cite{Oliver}.

The availability of 2D crystals allows for novel stacked heterostructures with strong proximity effects. Electronic modulation due to such substrates has been studied in graphene, such as hBN or twisting of another graphene layer \cite{Moiré-bands-in-twisted-double-layer-graphene,JarrilloHerrero,mori2,bilayer,valley,Gorbachev448}.
Lattice commensurability in these heterostructures depends on factors such as isotropic expansion, relative sliding between layers, and relative twists \cite{mori1,mori2,mori3}.

An interesting family of 2D crystals, transition metal dichalcogenides (TMD) can be used
as substrates for graphene \cite{vanderWaalsH,mog,mogrf,w2,mogrf2}.  Monolayer semiconductor TMD such as MoS$_2$ and WS$_2$ have a direct band gap
and honeycomb crystal structure \cite{mos2}.
The bands near the Fermi energy are formed predominantly from $d_{z^2}$, $ d_{xy}$ and $d_{x^2-y^2}$ orbitals
of the metal atom \cite{mos23}, with slight admixture from the $p$-orbitals of the chalcogen.  The SOC in the valence bands
is much larger than in the conduction bands, with strength that varies with the transition metal \cite{mos24}.
Successful growth of graphene on MoS$_2$ and WS$_2$ has been demonstrated experimentally \cite{G-MOS2-exp,w2,g-ws2}.
First principles calculations on some of these systems have proved challenging \cite{mogrf,w2,mogrf2}, with reported results that differ qualitatively
and quantitatively.

Motivated by these works, we study the topological properties of the minimal time reversal invariant effective model of
graphene that incorporates geometrical and orbital perturbation effects expected in these systems.
We focus on the Berry curvature and associated valley Chern number
and identify different quantum phases that may appear as Hamiltonian parameters vary.
The phase diagram shows that under the right conditions, it is possible to achieve band inversion of spin-split bands in graphene that acquire
interesting characteristics from the proximal TMD layer.  We further identify that a relative voltage difference between the
graphene and TMD layer (as obtained by an applied external field) can derive a transition between two topologically inequivalent phases, separated by a semimetallic phase.

The reduction of the spatial symmetries of graphene, and the enhancement of SOC in these systems results in the generation
of spin resolved gaps at the Dirac point. The interplay between sublattice symmetry breaking and enhanced SOC parameters
determines the size and topological nature of the gaps in the system. As the gaps are dominated by the SOC,
the system becomes a quantum spin Hall insulator, with symmetry protected edge states.  In contrast, when the gaps
are dominated by the sublattice staggered symmetry, the system becomes a valley Hall insulator. 

Identification of experimentally relevant parameters is carried out utilizing a tight binding formalism with appropriate
graphene and TMD characteristics.
Different lattice orientations and relative layer displacement are also found to exhibit such phase transition, with shifts in the values of external field at which it occurs.

\paragraph{Effective model and characteristics.}
The proximity of the TMD monolayer to graphene breaks inversion symmetry, which allows for the presence of Rashba SOC,
in addition to sublattice asymmetry terms in the effective Hamiltonian at low energies \cite{mahmoud-PRL}.
A minimal low energy model will include terms that respect
time reversal symmetry \cite{mogrf2,mahmoud-PRB}, and arise due to the symmetries in the TMD states,
$
\mathcal{H}_{\rm{eff}}=\mathcal{H}_{0}+\mathcal{H}_{\Delta}+\mathcal{H}_{S_1}+\mathcal{H}_{S_2}+\mathcal{H}_R
$,
with
\begin{equation}\label{eff-term}
\begin{split}
 \mathcal{H}_{0}&=\hbar v_F \left( \tau_z\sigma_{x}s_0p_{x}+\tau_0\sigma_{y}s_0p_{y}\right)\\
 \mathcal{H}_{\Delta}&=\Delta s_0 \sigma_z \tau_0\\
 \mathcal{H}_{S_1}&=S_1 \tau_z\sigma_zs_z\\
 \mathcal{H}_{S_2}&=S_2\tau_z\sigma_0s_z\\
 \mathcal{H}_R&= R (\tau_z\sigma_xs_y-\tau_0\sigma_ys_x)\\
\end{split}
\end{equation}
where $ \sigma_i$, $ \tau_i$, and $s_i$ are $2\times2$ Pauli matrices with $i={0,x,y,z}$, (where 0 is used for the unit matrix) operating
on different degrees of freedom. $\sigma_i$ acts on the pseudospin sublattice space  (A,B), $ \tau_i $ on the K, K$ ' $ valley space,
and $ s_i $ on the spin degree of freedom \cite{mahmoud-PRL}.
We use the `standard' basis $\Psi^T = (\Psi^T_K, \Psi^T_{K'})$, with
$\Psi_{K,K'}^T= (A\uparrow,B\uparrow,A\downarrow,B\downarrow)_{K,K'}$, and {$\mathcal{H}_0$}
describes pristine graphene at low energy \cite{qshe}.
The parameters $v_F,\Delta,S_1,S_2$, and $R$ are constants of the model, to be obtained from DFT or tight-binding calculations
(as we describe below).  They would naturally be expected to depend on the microscopic details of the system, such as orientation
and relative displacements of the monolayers, as well as on applied electric fields.
As we will see below, it is such dependence that may give rise to interesting phases.

Different terms in the effective model play interesting physical roles.
$ \mathcal{H}_{\Delta}$ characterizes the (staggered) sublattice asymmetry in the graphene A and B atoms, as one expects
from the proximity to the TMD monolayer; this term is well-known to open gaps in the otherwise linear dispersion of $\mathcal{H}_{0}$, and create sizable topological-valley currents in graphene-hBN superlattices \cite{Gorbachev448,mahmoud-PRB}.
The intrinsic SOC term, $\mathcal{H}_{S_1}$, opens a spin gap in the bulk structure with opposite signs at the K and K$'$ valleys,
while preserving spatial symmetries of the hexagonal lattice.
Finally, as mirror symmetry ($z\rightarrow -z$) is broken in the presence of the TMD substrate, the dynamics is expected to contain a
Rashba effective Hamiltonian $\mathcal{H}_R$ \cite{qshe}, and a diagonal SOC term
$ \mathcal{H}_{S_2}$.
Although a valley mixing term is possible in principle, we find it to be essentially null in all our calculations.

\begin{figure}
	\includegraphics[width=1.0\linewidth]{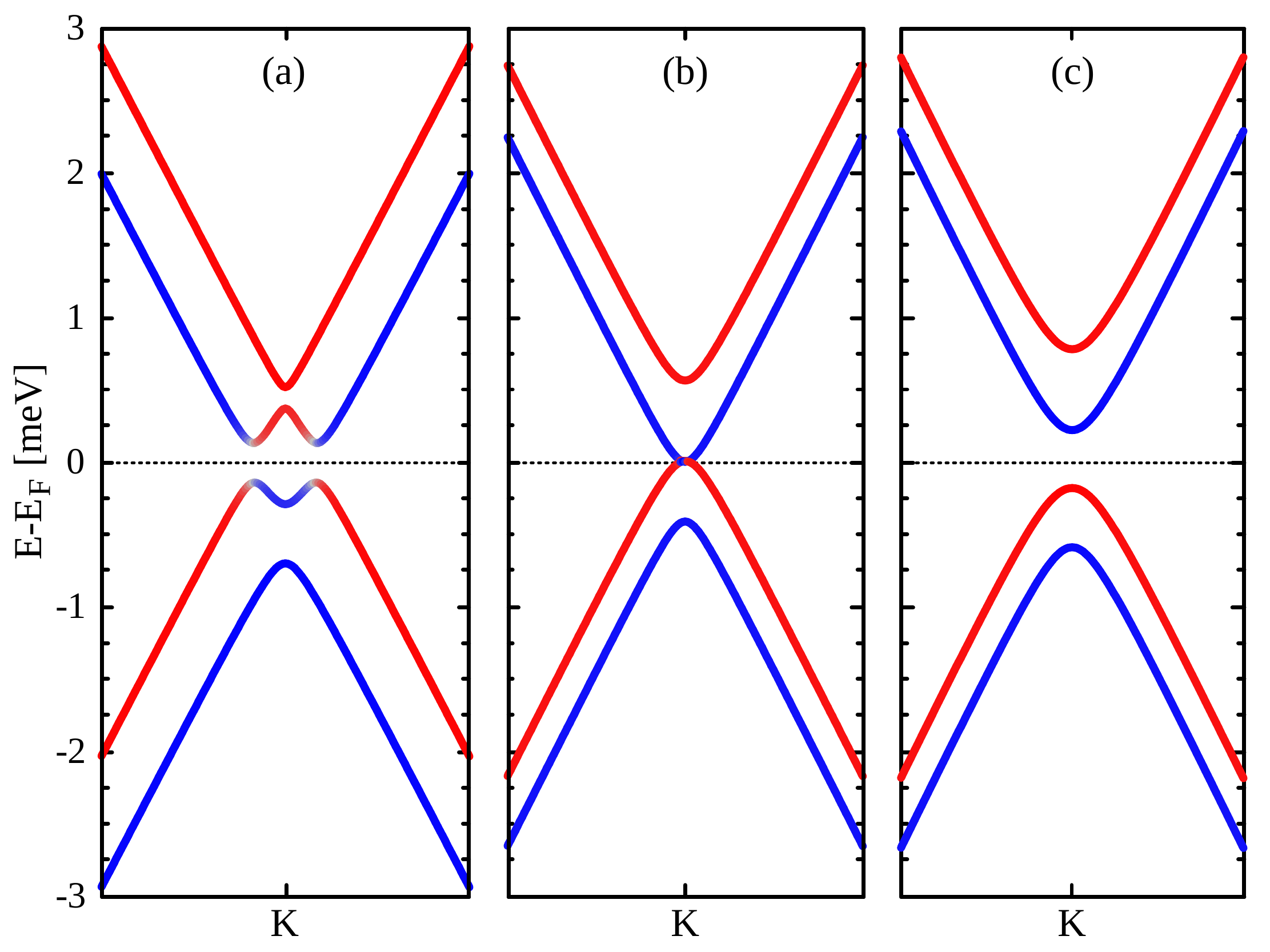}
	\caption{(Color online) Typical band structure of effective model near the K valley.  Left panel shows an `inverted band' regime, with strong spin mixing of the different states, as indicated by the red/blue shading, and typical of $|S_1 + S_2| > \Delta$.  Middle panel shows a spin split semimetallic phase, while right panel shows a 'direct band' regime where a finite bulk gap develops with nearly full spin polarization, obtained when $|S_1 + S_2| < \Delta$.}
	\label{fig1}
\end{figure}

Typical band structures for this Hamiltonian are shown in Fig.\ \ref{fig1}. The left panel illustrates an `inverted band' regime, evident in the local dispersion around each of the valleys near the graphene neutrality point, produced by the anticrossing of bands with opposite
spins and due to the presence of the Rashba term.  The middle panel shows a transition point, where the gap has closed and exhibits a dispersion with nearly full spin polarization.  The right panel shows a `direct band' regime with a simple parabolic dispersion for each of the two spin projections.  As we will see below, the inverted band regime is achieved whenever $|S_1 + S_2| > \Delta$, while the direct band regime is achieved in the opposite case. 

\begin{figure}
		\includegraphics[width=1.0\linewidth]{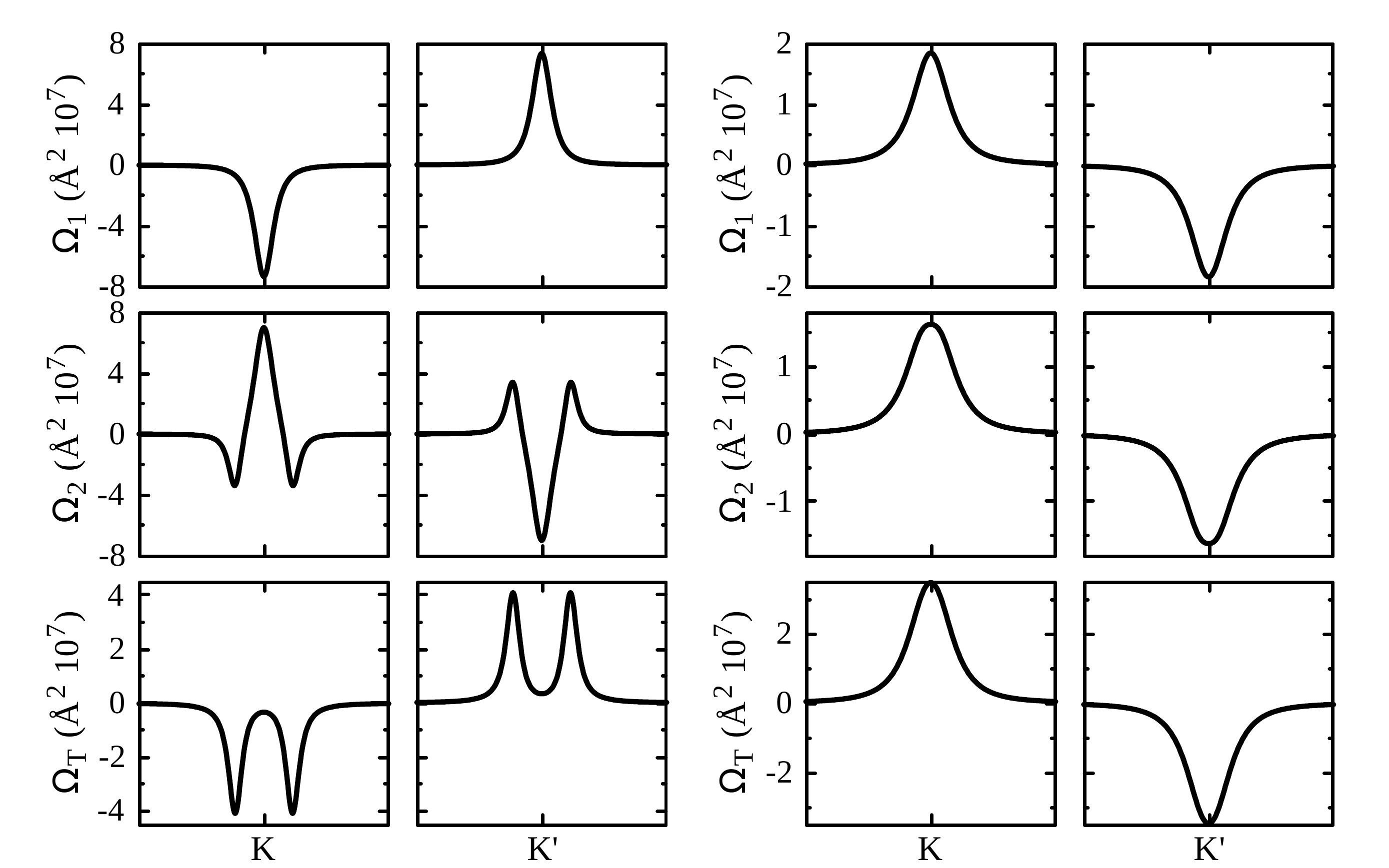}
\caption{Berry curvature $\Omega_n$ at K and K$'$ valleys for both inverted and direct band gap regimes. Left two columns show results
for the inverted band regime corresponding to Fig.\ \ref{fig1}a.  Right two columns are for the direct band regime corresponding to Fig.\ \ref{fig1}c.
Upper (middle) plots describe Berry curvature of the lowest (highest) energy valence bands in Fig.\ref{fig1}, $n=1(2)$. Lower plots show
the total valence band Berry curvature, $\Omega_T = \Omega_1 + \Omega_2$. The different Berry curvature distribution between K and K$'$
is evident in both cases. }
	\label{fig2}
\end{figure}

One can analyze the topological features of the states described by the effective Hamiltonian (\ref{eff-term}) by calculating the Berry
curvature $\Omega_{n}(\boldsymbol{k})$ and Chern number per valley of the occupied bands using \cite{berry}
\begin{eqnarray}\label{berry}
\Omega_{n}(\boldsymbol{k})&=&-\sum_{n'\neq n}\frac{2 {\rm Im}\langle\Psi_{n'\boldsymbol{k}}| v_x |\Psi_{n\boldsymbol{k}}\rangle\langle\Psi_{n\boldsymbol{k}}| v_y|\Psi_{n'\boldsymbol{k}}\rangle}{(\epsilon_n-\epsilon_{n'})^2}, \nonumber \\
\mathcal{C}_n&=&\frac{1}{2\pi}\int dk_x dk_y \Omega_n(k_x,k_y),
\end{eqnarray}
where $n$ is the band number, and ${v}_x ({v}_y)$ is the velocity operator along the $x(y)$ direction \cite{graphene-Chern}.
Figure \ref{fig2} shows Berry curvature for the two lowest energy (valence) bands,
and total curvature near each of the
K and K$ ' $ valleys, in two different parameter regimes.   Notice plots for each band obey
$\Omega$(K-valley$) = -\Omega$(K$ ' $-valley), as required by time reversal symmetry \cite{berry}. 
The left two columns in Fig.\ \ref{fig2}, for the inverted band regime, exhibit a non-monotonic $k$-dependence for the curvature
in each valley, with inversion at each K point, $\Omega_1(0) \simeq -\Omega_2(0)$, so that the total valley curvature is nearly null.
In contrast, the right two columns for the direct band regime show the same curvature for both bands in each valley.
The non-vanishing Berry curvature in each valley may give rise to interesting edge states in systems with borders, as seen in graphene ribbons and TMD flake edges \cite{ribons2,Li2011,CarlosEdges}.

The total Chern index at each valley yields $\mathcal{C}_{\rm K}=-\mathcal{C}_{\rm K'}$ for all
parameter values, so that the overall Chern number vanishes, as expected for systems protected by time reversal symmetry \cite{theory,berry}.
However, the spin splitting and mixing of the two valence bands in different regimes results in $\mathcal{C}_{\rm K}=\pm 1$,
with an overall sign change across the semimetallic phase transition where the gap closes.
In a system with zero Rashba term ($R=0$), the Chern number can be shown to yield $\mathcal{C} =  {\rm sgn} (\Delta \tau_z + S_1 s_z)$,
indicating the competition between the intrinsic SOC and staggered perturbations.
Although an analytic expression for the Chern number is not feasible in general, numerical evaluation for different parameter regimes reveals the important
roles of both $S_1$ and $S_2$, as well as $R$, on determining the topological features of the system.  We will return to this below in detail.

\begin{figure}
	\centering
	\begin{minipage}{0.12\textwidth}
		\includegraphics[width=1.0\linewidth]{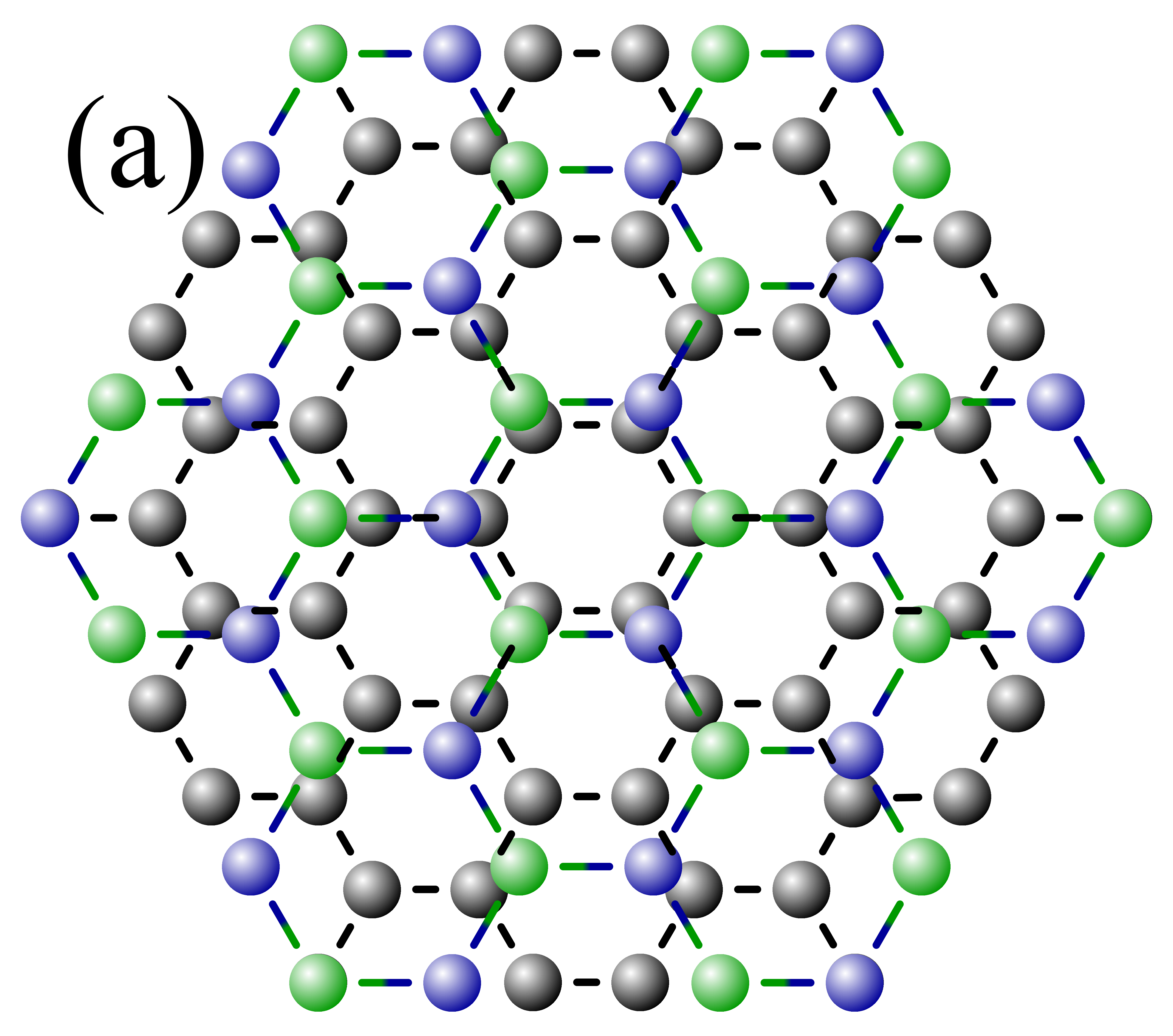}
		\vspace{-3ex}
		\label{fig3-a}
		\includegraphics[width=1.0\linewidth]{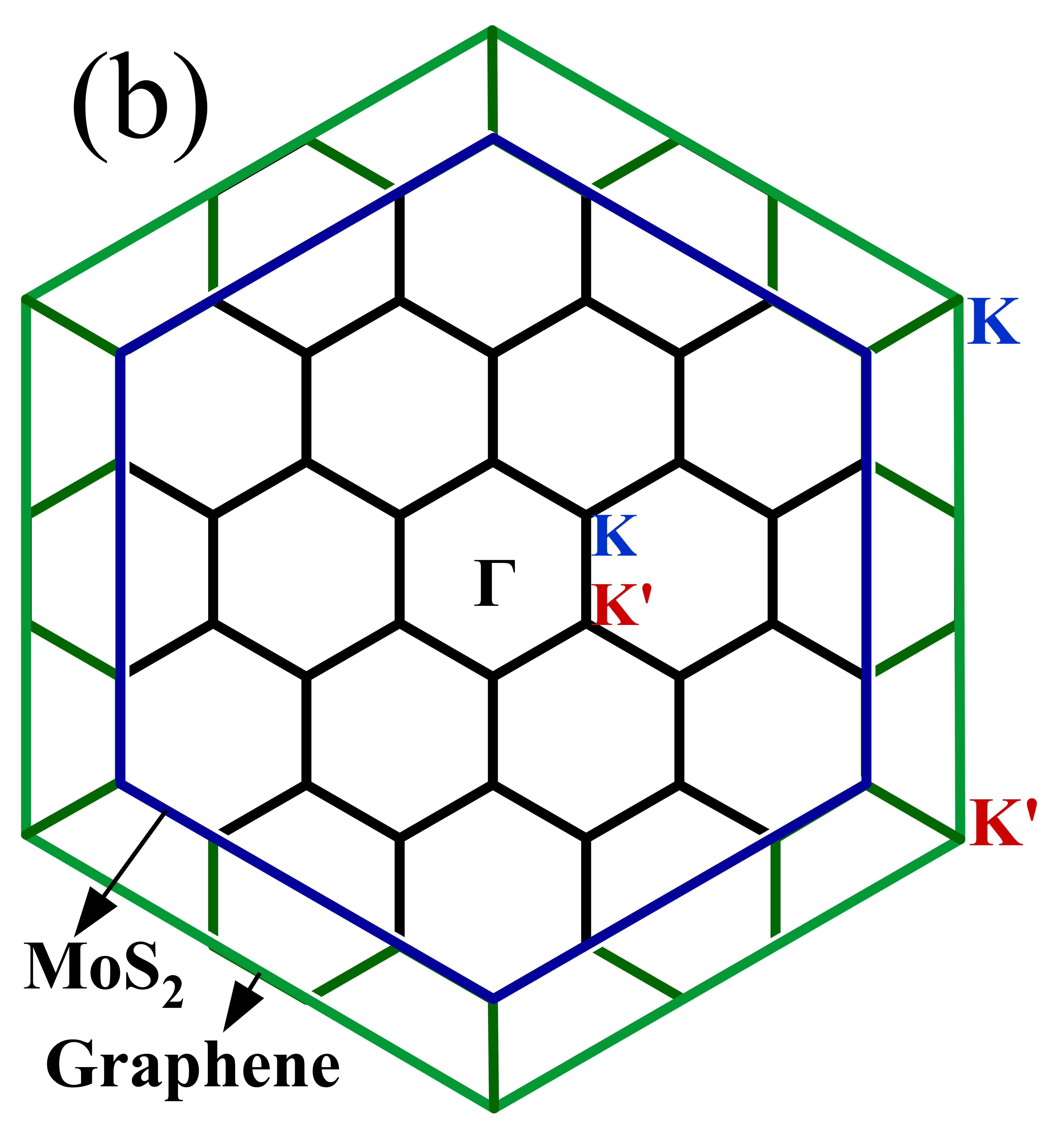}
		\label{fig3-b}
	\end{minipage}
	\begin{minipage}{0.33\textwidth}
		\includegraphics[width=1.0\linewidth]{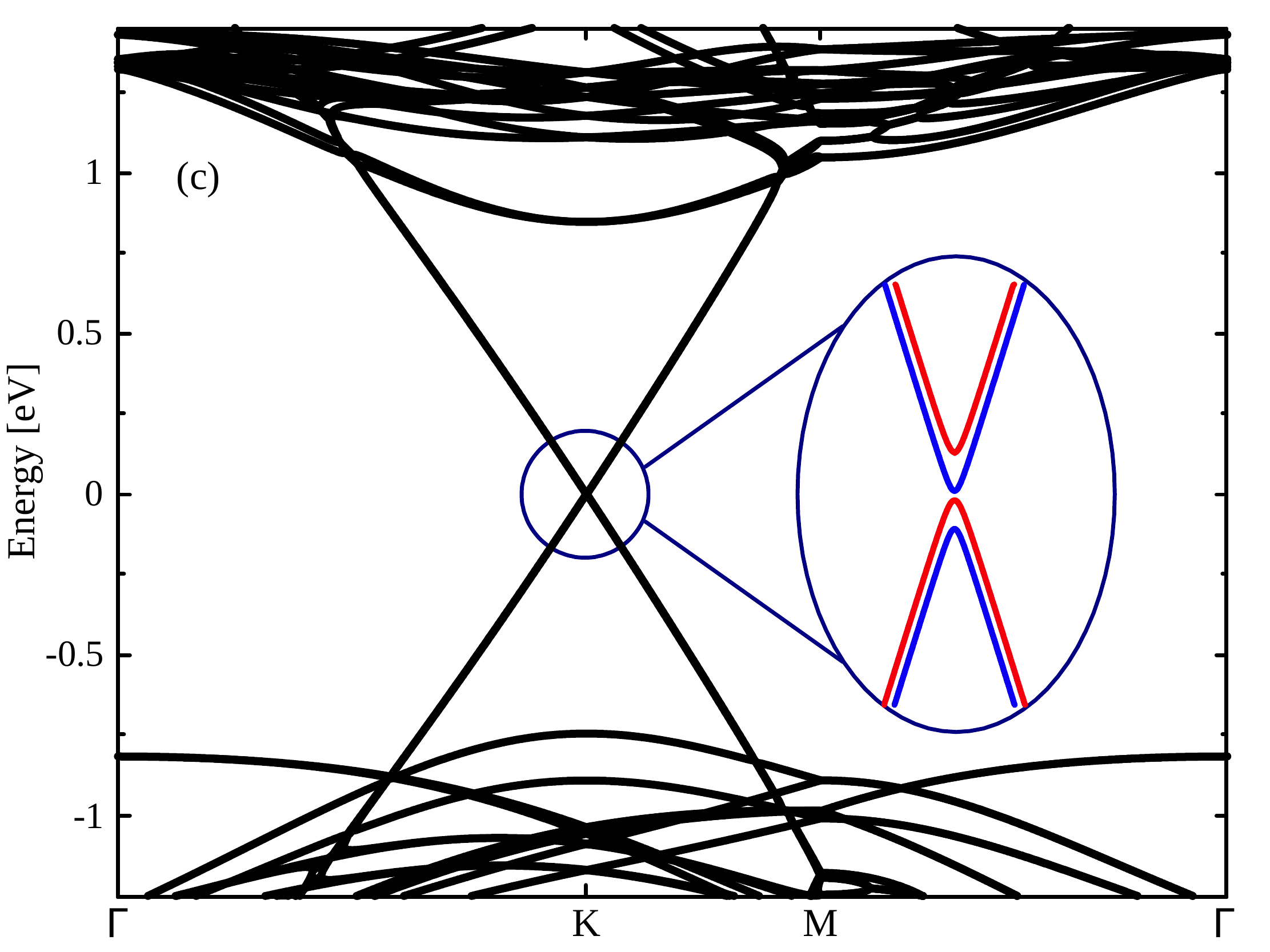}
	\label{fig3-c}
	\end{minipage}
	\caption{(Color online) Graphene-MoS$_2$ heterostructure. (a) Top view of bilayer structure supercell in real space.
Black circles are A and B atoms of graphene, while blue (green) are Mo (S$_2$) atoms. (b) Brillouin zones of the reciprocal lattices:
First BZ for a monolayer of graphene and MoS$_2$ with position of K and K' valleys. Upon folding onto the heterostructure reciprocal lattice,
corner valleys from both layers are mapped onto the same point. (c) Band dispersion of graphene-MoS$_2$ along high symmetry lines
$\Gamma$-K-M-$\Gamma$. Inset: Zoom near K valley shows graphene bands appear gapped and spin polarized due to
proximity to MoS$_2$. Blue (red) bands are for spin up (down) states.}
	\label{fig3}
\end{figure}

\paragraph{Tight binding model.}
To study the relevant effective model dependence on microscopic details of the graphene-TMD heterostructure, we have
implemented a tight-binding model of the structure.
For specificity, we focus on graphene and MoS$_2$, with lattice constants 2.46 and 3.11 \si\angstrom, respectively.
A superlattice of 5$ \times $5 graphene unit cells and 4$ \times $4
MoS$_2$ results in a nearly commensurate moire pattern with a small residual strain ($\sim$1.1 $\%$), as seen in Fig.\ \ref{fig3}a.

The primitive lattice vectors in such superlattice are connected by a linear transformation \cite{mori1},
$ 
\left(
\textbf{a}_{G_1},
\textbf{a}_{G_2}
\right)^T
=\doubleunderline{M}
\left(
\textbf{a}_{\rm {Mo}_1},
\textbf{a}_{\rm{Mo}_2}
\right)^T
$, 
where $ \textbf{a}_{x_i}$ are the primitive vectors in each layer ($x$ = graphene and MoS$_2$).
The superlattice has primitive lattice vectors given by
$
\left(
\textbf{R}_{1},
\textbf{R}_{2}
\right)^T
={[\doubleunderline{1}-\doubleunderline{M}]^{-1}\doubleunderline{M}} 
\left(
\textbf{a}_{\rm{Mo}_1},
\textbf{a}_{\rm{Mo}_2}
\right)^T
$; 
here $\doubleunderline{M}= {\rm diag}(\frac{4}{5},\frac{4}{5})$.
The reduced Brillouin zone has similar features to that of graphene, with valleys at
$\textbf{K}=\frac{2\pi}{a_\alpha}\left( \frac{1}{\sqrt{3}},\frac{1}{3}\right)$,
$\textbf{K}' =\frac{2\pi}{a_\alpha}\left( \frac{1}{\sqrt{3}},\frac{-1}{3}\right)$, $a_\alpha=5 a_G = 4 a_{\rm{Mo}}$,
which fold the corresponding valleys of graphene and MoS$_2$ onto the same points; see Fig.\ \ref{fig3}b.

The tight-binding formalism couples nearest neighbors $\langle ij \rangle$ in an optimal basis where MoS$_2$
is represented by three orbitals, $d_{z^2}$, $d_{xy}$ and $d_{x^2-y^2}$\cite{mos23},
\begin{equation}\label{mo-h}
\begin{split}
\mathcal{H}_{{\rm Mo}}=\sum_{i  \nu s}\epsilon_{\nu s} \alpha^{\dagger}_{i\nu s}\alpha_{i\nu  s}
+\sum_{\langle ij\rangle  \nu\mu s} t_{i\nu,j\mu} \alpha_{i\nu s}^{\dagger}\alpha_{j\mu s}+h.c.
\end{split}
\end{equation}
$\epsilon_{\nu s}$ considers the on-site energies of Mo-atom $i$, orbital $\nu$, and spin $s$, while $t_{i\nu,j\mu}$ describes nearest
neighbor hopping between Mo orbitals. The SOC in MoS$_2$ is introduced via atomic contributions \cite{mos23}.
For graphene we adopt the usual $p_z$-orbital representation with two-atom basis \cite{graphene,Suppl}.
The substrate generates an electric field normal to the layer, causing
a Rashba SOC term \cite{qshe},
$ 
{\mathcal H}_{R}= i t_R\sum_{\langle ij\rangle \alpha \beta} \hat{{z}}\cdot({\bf s}_{\alpha\beta}\times {\bf d}^{\circ}_{ij}) c^{\dagger}_{i\alpha}c_{j\beta}$, 
where $\alpha,\beta$ describe spin up and down states, and ${\textit{\bf d}}^{\circ}_{ij}$ is the unit vector that connects neighbor atoms A and B.
Although the Rashba
interaction is weak in graphene ($t_R=0.067$meV \cite{G-Rash}), it is an important term that breaks inversion
symmetry.

The interlayer coupling between graphene $ p_z $ orbital and MoS$_2$ $ d$-orbitals is given by
\begin{equation}\label{T-B-h}
\begin{split}
\mathcal{H}=\sum_{\langle ij\rangle,\nu\sigma} t^{\nu}_{i,j} {c}_{i,\sigma}^{\dagger}{\alpha}_{j\nu,\sigma}+h.c.
\end{split}
\end{equation}
 We take coupling to nearest neighbors across layers, where
$ 
t^{\nu}_{i,j}=t_{\nu} \text{ exp}\left[ -\arrowvert \textbf{r}_{m,i}-\textbf{r}_{g,j}\arrowvert/\eta\right] $
is parameterized by the distance connecting atoms in both layers, $\arrowvert \textbf{r}_{m,i}-\textbf{r}_{g,j}\arrowvert$,
normalized to a constant $\eta=5a_g$; $t_{\nu}$ describes the coupling between $p_z$ and $d$-orbitals
using a Slater-Koster approach \cite{slater}. The couplings depend naturally on the orbitals involved,
which results in $ t_{z^2} $ being larger than $ t_{xy} $ and  $t_{x^2-y^2} $, due to a higher overlap.
The parameters used are described in the supplement \cite{Suppl}, although the detailed values
do not affect the main conclusions nor qualitative behavior, providing only an overall scaling of parameter ranges.

The tight-binding model considers the possibility of a difference in electronegativity between the two layered materials creating a relative shift
of their neutrality points.  This polarization shift could further be thought to arise from an applied voltage between the layers, as it would be
possible (in principle) to apply if the graphene-MoS$_2$ structure is placed between capacitor plates.  We explore the consequences
of such relative voltage on the effective band structure on graphene, assuming that the other parameters (hopping integrals and
lattice constants) remain unchanged with voltage.  One could obtain the appropriate parameters from first principles calculations,
although the van der Waals nature of the bonding between layers, as well as the rather fine-scale of the relevant features make those
calculations quite challenging \cite{w2,mogrf}.
Results for a nearly zero relative shift of the neutrality points are in Fig.\ \ref{fig3}c, which show how the low energy spectrum
exhibits a finite gap for fully spin polarized bands.

\begin{figure}
	\includegraphics[width=0.95\linewidth]{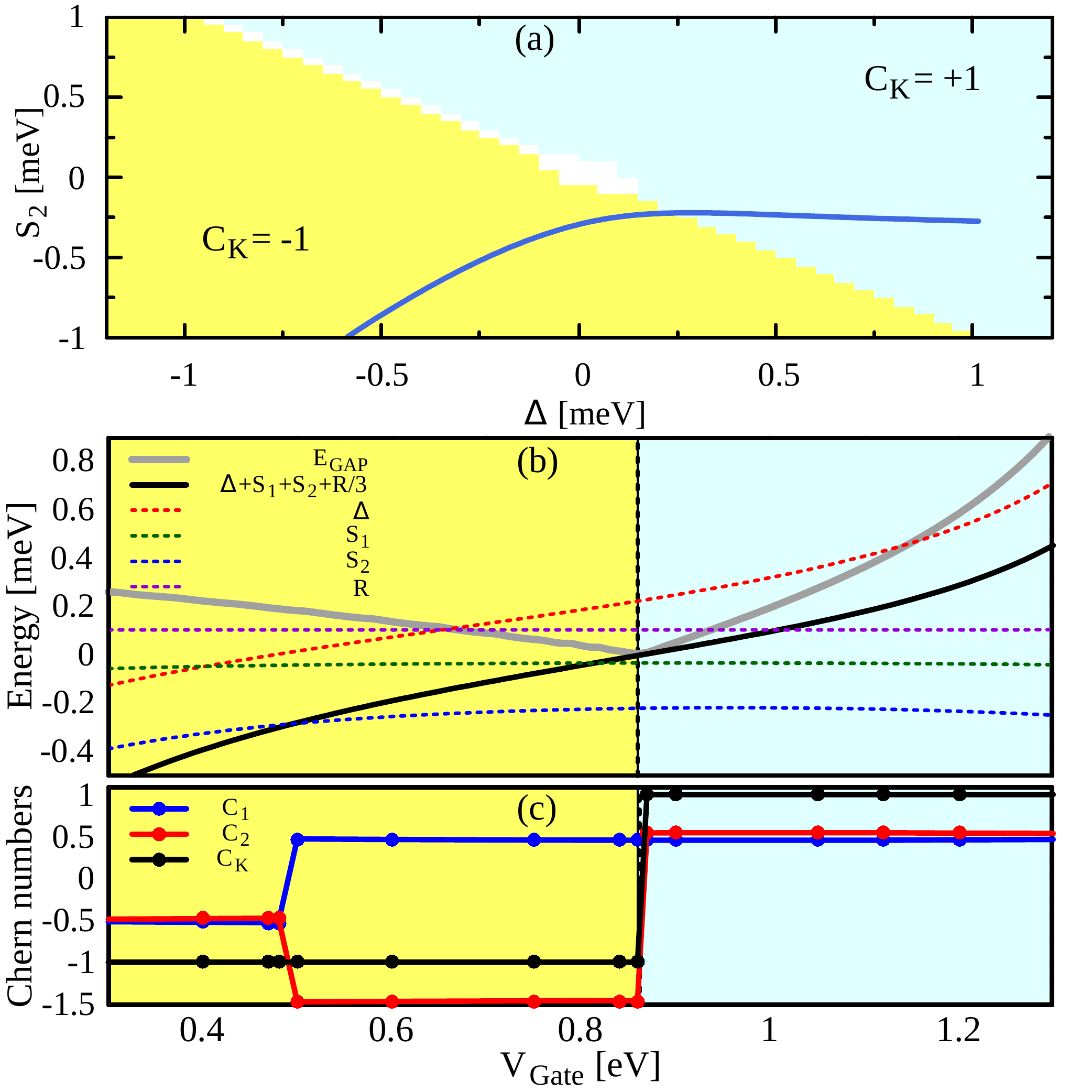}
\caption{(Color online) 
(a) Phase diagram for Graphene-TMD system in Eq. \ref{eff-term} in the $S_{2}$-$\Delta$ projection with $R=0.1m$eV, and $S_{1} \in [-0.16, 0.16 ]m$eV. Trivial insulating phase in blue $C_{K}=1$, and mass inverted phase in yellow $C_{K}=-1$, divided by the semimetallic phase, white curve. Blue line shows the line cut for graphene-MoS$_{2}$ system as a function of V$_{Gate}$. (b) Gate voltage dependence of effective Hamiltonian parameters used to fit the tight-binding band structure results. Black line shows evolution of band gap, which closes at $V_{Gate} = 0.86$ eV.  For $V_{Gate}< 0.86$ eV, in the inverted band regime, spin orbit contribution dominates over the staggered term, given approximately
by $\Delta+S_1+S_2 <0$, slightly shifted by Rashba term. In the opposite regime, staggered potential
dominates and creates a trivial band gap.(c) Chern numbers for K valley, as in (a).
Switch near $V_{Gate}=0.5$ eV is due to a band crossing at the K point,
while jump at  $V_{Gate}=0.86$ eV indicates gap closing that separates inverted mass regime from direct band regime. }
	\label{fig4}
\end{figure}

As the relative voltage between layers is varied, the tight-binding spectrum shows a low-energy band structure similar to that of
the effective model, Fig.\ \ref{fig1}.
We have carried out a systematic fit of the low energy dispersion with the model
parameters in Eq.\ \ref{eff-term}, as the voltage changes.
The fits are excellent (to less than one percent) up to an energy 0.3 eV away from the graphene Dirac point.
The effective model describes not only the low-energy band dispersion, but also the full spin and pseudospin structure of the
states, illustrating the generality of the model \cite{Suppl}.
Fit parameters vary smoothly
with gate voltage, as shown in Fig.\ \ref{fig4}b; we assign $V_{Gate}=0$ when
graphene Dirac point is 20 meV higher than the top valence band in MoS$_2$, while a zero relative shift of their neutrality points is
at $V_{Gate} \simeq 0.9 \si\eV$.
The staggered potential $\Delta$ increases smoothly with voltage, while $R$, $S_1$, and $S_2$ vary much less \footnote{We have
kept $t_R$ constant in the tight binding calculations, although a gate voltage dependence is likely.  This would be expected to reduce
the gate voltage required to close the gap.}.
Most importantly, we see that the gap closes near the gate voltage where the sum $\Delta+S_1+S_2+R/3 \simeq 0$.
Figure \ref{fig4}a and b also show the valley Chern number jumps by $2\pi$ at
the closing of the gap, as anticipated from the discussion above, although here the competition involves $\Delta$ and all three
SOC coefficients.
We emphasize that the closing of the gap and corresponding phase transition from an inverted
mass to a direct gap regime is rather generic, and as suggested by these calculations, accessible experimentally.

It is important to notice that the model studied is one particular example of a large class of Hamiltonians that describe systems that possess similar
symmetry properties with different parameters. More examples are
graphene and other TMD heterostructures, with some quantitative differences.
In graphene-WS$_2$, the inverted band phase exists over a wider range of gate voltage, with gap closing
at $V_{\rm Gate} = 1.2$ eV.  This larger voltage can be understood as
arising from the larger SOC in WS$_2$, nearly three times stronger than in MoS$_2$ \cite{mos2}. We also explored structures with a relative shift of the lattices, or possible rotations of the two layers involved.
We find that gaps open generically near the graphene neutrality point, with regimes of inverted masses, at times over only narrow regions of gate voltage.  This suggests that in a macroscopic sample with a distribution
of strains, one may expect
variation of the effective Hamiltonian parameters over long range scales. This may produce `edge states' separating different
regions in the 2D bulk with different topological features, resulting interesting effects even away from the sample edges.

 The topology of the inverted mass regime in the structure suggests that interesting edge states would exist at boundaries,
as discussed in the past \cite{berry}. In fact, zigzag edge graphene nanoribbons based on the effective Hamiltonian show different regimes.  A quantum spin Hall effect is seen for the mass inverted bands, while a valley Hall effect is present in the direct band regime.

\paragraph{Conclusions.}
We have built and studied a heterostructure of graphene deposited on a monolayer of transition metal dichalcogenides (TMD) in order to explore  proximity effects. An effective Hamiltonian
with possible perturbations that preserve time reversal is able to faithfully reproduce the results from tight binding calculations of the structure near the graphene Dirac points.
The proximity of the TMD results in sizable spin orbit coupling imparted onto graphene, in a degree proportional
to the intrinsic SOC in the TMD.  This strong effect is found to compete with the staggered potential also introduced, resulting
in different regimes where the heterostructure changes phase from an inverted mass band structure, with possible quantum spin Hall effect and the consequent spin filtered edge states, to a direct band structure with possible valley Hall effect and the appearance of valley currents.  These phases could in principle be controlled by a relative gate voltage between the layers, and may even be present throughout the 2D bulk, as strains fields would affect the relevant phases present.

\emph{Acknowledgments}. We acknowledge support from NSF-DMR 1508325, the Saudi Arabian Cultural Mission to the US for a Graduate Scholarship, and hospitality of the Aspen Center for Physics, supported by NSF-PHY 1066293.
\bibliography{cites}

\begin{thebibliography}{38}%
\makeatletter
\providecommand \@ifxundefined [1]{%
 \@ifx{#1\undefined}
}%
\providecommand \@ifnum [1]{%
 \ifnum #1\expandafter \@firstoftwo
 \else \expandafter \@secondoftwo
 \fi
}%
\providecommand \@ifx [1]{%
 \ifx #1\expandafter \@firstoftwo
 \else \expandafter \@secondoftwo
 \fi
}%
\providecommand \natexlab [1]{#1}%
\providecommand \enquote  [1]{``#1''}%
\providecommand \bibnamefont  [1]{#1}%
\providecommand \bibfnamefont [1]{#1}%
\providecommand \citenamefont [1]{#1}%
\providecommand \href@noop [0]{\@secondoftwo}%
\providecommand \href [0]{\begingroup \@sanitize@url \@href}%
\providecommand \@href[1]{\@@startlink{#1}\@@href}%
\providecommand \@@href[1]{\endgroup#1\@@endlink}%
\providecommand \@sanitize@url [0]{\catcode `\\12\catcode `\$12\catcode
  `\&12\catcode `\#12\catcode `\^12\catcode `\_12\catcode `\%12\relax}%
\providecommand \@@startlink[1]{}%
\providecommand \@@endlink[0]{}%
\providecommand \url  [0]{\begingroup\@sanitize@url \@url }%
\providecommand \@url [1]{\endgroup\@href {#1}{\urlprefix }}%
\providecommand \urlprefix  [0]{URL }%
\providecommand \Eprint [0]{\href }%
\providecommand \doibase [0]{http://dx.doi.org/}%
\providecommand \selectlanguage [0]{\@gobble}%
\providecommand \bibinfo  [0]{\@secondoftwo}%
\providecommand \bibfield  [0]{\@secondoftwo}%
\providecommand \translation [1]{[#1]}%
\providecommand \BibitemOpen [0]{}%
\providecommand \bibitemStop [0]{}%
\providecommand \bibitemNoStop [0]{.\EOS\space}%
\providecommand \EOS [0]{\spacefactor3000\relax}%
\providecommand \BibitemShut  [1]{\csname bibitem#1\endcsname}%
\let\auto@bib@innerbib\@empty
\bibitem [{\citenamefont {Castro~Neto}\ \emph {et~al.}(2009)\citenamefont
  {Castro~Neto}, \citenamefont {Guinea}, \citenamefont {Peres}, \citenamefont
  {Novoselov},\ and\ \citenamefont {Geim}}]{graphene}%
  \BibitemOpen
  \bibfield  {author} {\bibinfo {author} {\bibfnamefont {A.~H.}\ \bibnamefont
  {Castro~Neto}}, \bibinfo {author} {\bibfnamefont {F.}~\bibnamefont {Guinea}},
  \bibinfo {author} {\bibfnamefont {N.~M.~R.}\ \bibnamefont {Peres}}, \bibinfo
  {author} {\bibfnamefont {K.~S.}\ \bibnamefont {Novoselov}}, \ and\ \bibinfo
  {author} {\bibfnamefont {A.~K.}\ \bibnamefont {Geim}},\ }\href {\doibase
  10.1103/RevModPhys.81.109} {\bibfield  {journal} {\bibinfo  {journal} {Rev.
  Mod. Phys.}\ }\textbf {\bibinfo {volume} {81}},\ \bibinfo {pages} {109}
  (\bibinfo {year} {2009})}\BibitemShut {NoStop}%
\bibitem [{\citenamefont {Kane}\ and\ \citenamefont {Mele}(2005)}]{qshe}%
  \BibitemOpen
  \bibfield  {author} {\bibinfo {author} {\bibfnamefont {C.~L.}\ \bibnamefont
  {Kane}}\ and\ \bibinfo {author} {\bibfnamefont {E.~J.}\ \bibnamefont
  {Mele}},\ }\href {\doibase 10.1103/PhysRevLett.95.226801} {\bibfield
  {journal} {\bibinfo  {journal} {Phys. Rev. Lett.}\ }\textbf {\bibinfo
  {volume} {95}},\ \bibinfo {pages} {226801} (\bibinfo {year}
  {2005})}\BibitemShut {NoStop}%
\bibitem [{\citenamefont {Gmitra}\ \emph {et~al.}(2009)\citenamefont {Gmitra},
  \citenamefont {Konschuh}, \citenamefont {Ertler}, \citenamefont
  {Ambrosch-Draxl},\ and\ \citenamefont {Fabian}}]{gdft}%
  \BibitemOpen
  \bibfield  {author} {\bibinfo {author} {\bibfnamefont {M.}~\bibnamefont
  {Gmitra}}, \bibinfo {author} {\bibfnamefont {S.}~\bibnamefont {Konschuh}},
  \bibinfo {author} {\bibfnamefont {C.}~\bibnamefont {Ertler}}, \bibinfo
  {author} {\bibfnamefont {C.}~\bibnamefont {Ambrosch-Draxl}}, \ and\ \bibinfo
  {author} {\bibfnamefont {J.}~\bibnamefont {Fabian}},\ }\href {\doibase
  10.1103/PhysRevB.80.235431} {\bibfield  {journal} {\bibinfo  {journal} {Phys.
  Rev. B}\ }\textbf {\bibinfo {volume} {80}},\ \bibinfo {pages} {235431}
  (\bibinfo {year} {2009})}\BibitemShut {NoStop}%
\bibitem [{\citenamefont {Pesin}\ and\ \citenamefont
  {MacDonald}(2012)}]{spintronics}%
  \BibitemOpen
  \bibfield  {author} {\bibinfo {author} {\bibfnamefont {D.}~\bibnamefont
  {Pesin}}\ and\ \bibinfo {author} {\bibfnamefont {A.~H.}\ \bibnamefont
  {MacDonald}},\ }\href@noop {} {\bibfield  {journal} {\bibinfo  {journal}
  {Nature Materials}\ }\textbf {\bibinfo {volume} {11}},\ \bibinfo {pages}
  {409} (\bibinfo {year} {2012})}\BibitemShut {NoStop}%
\bibitem [{\citenamefont {Avsar}\ \emph {et~al.}(2015)\citenamefont {Avsar},
  \citenamefont {Lee}, \citenamefont {Koon},\ and\ \citenamefont
  {\"O�zyilmaz}}]{fluor}%
  \BibitemOpen
  \bibfield  {author} {\bibinfo {author} {\bibfnamefont {A.}~\bibnamefont
  {Avsar}}, \bibinfo {author} {\bibfnamefont {J.~H.}\ \bibnamefont {Lee}},
  \bibinfo {author} {\bibfnamefont {G.~K.~W.}\ \bibnamefont {Koon}}, \ and\
  \bibinfo {author} {\bibfnamefont {B.}~\bibnamefont {\"O�zyilmaz}},\ }\href
  {http://stacks.iop.org/2053-1583/2/i=4/a=044009} {\bibfield  {journal}
  {\bibinfo  {journal} {2D Materials}\ }\textbf {\bibinfo {volume} {2}},\
  \bibinfo {pages} {044009} (\bibinfo {year} {2015})}\BibitemShut {NoStop}%
\bibitem [{\citenamefont {Weeks}\ \emph {et~al.}(2011)\citenamefont {Weeks},
  \citenamefont {Hu}, \citenamefont {Alicea}, \citenamefont {Franz},\ and\
  \citenamefont {Wu}}]{WU}%
  \BibitemOpen
  \bibfield  {author} {\bibinfo {author} {\bibfnamefont {C.}~\bibnamefont
  {Weeks}}, \bibinfo {author} {\bibfnamefont {J.}~\bibnamefont {Hu}}, \bibinfo
  {author} {\bibfnamefont {J.}~\bibnamefont {Alicea}}, \bibinfo {author}
  {\bibfnamefont {M.}~\bibnamefont {Franz}}, \ and\ \bibinfo {author}
  {\bibfnamefont {R.}~\bibnamefont {Wu}},\ }\href {\doibase
  10.1103/PhysRevX.1.021001} {\bibfield  {journal} {\bibinfo  {journal} {Phys.
  Rev. X}\ }\textbf {\bibinfo {volume} {1}},\ \bibinfo {pages} {021001}
  (\bibinfo {year} {2011})}\BibitemShut {NoStop}%
\bibitem [{\citenamefont {Giovannetti}\ \emph {et~al.}(2015)\citenamefont
  {Giovannetti}, \citenamefont {Capone}, \citenamefont {van~den Brink},\ and\
  \citenamefont {Ortix}}]{ind-g}%
  \BibitemOpen
  \bibfield  {author} {\bibinfo {author} {\bibfnamefont {G.}~\bibnamefont
  {Giovannetti}}, \bibinfo {author} {\bibfnamefont {M.}~\bibnamefont {Capone}},
  \bibinfo {author} {\bibfnamefont {J.}~\bibnamefont {van~den Brink}}, \ and\
  \bibinfo {author} {\bibfnamefont {C.}~\bibnamefont {Ortix}},\ }\href
  {\doibase 10.1103/PhysRevB.91.121417} {\bibfield  {journal} {\bibinfo
  {journal} {Phys. Rev. B}\ }\textbf {\bibinfo {volume} {91}},\ \bibinfo
  {pages} {121417} (\bibinfo {year} {2015})}\BibitemShut {NoStop}%
\bibitem [{\citenamefont {Marchenko}\ \emph {et~al.}(2012)\citenamefont
  {Marchenko}, \citenamefont {Varykhalov}, \citenamefont {Scholz},
  \citenamefont {Bihlmayer}, \citenamefont {Rashba}, \citenamefont {Rybkin},
  \citenamefont {Shikin},\ and\ \citenamefont {Rader}}]{Oliver}%
  \BibitemOpen
  \bibfield  {author} {\bibinfo {author} {\bibfnamefont {D.}~\bibnamefont
  {Marchenko}}, \bibinfo {author} {\bibfnamefont {A.}~\bibnamefont
  {Varykhalov}}, \bibinfo {author} {\bibfnamefont {M.}~\bibnamefont {Scholz}},
  \bibinfo {author} {\bibfnamefont {G.}~\bibnamefont {Bihlmayer}}, \bibinfo
  {author} {\bibfnamefont {E.}~\bibnamefont {Rashba}}, \bibinfo {author}
  {\bibfnamefont {A.}~\bibnamefont {Rybkin}}, \bibinfo {author} {\bibfnamefont
  {A.}~\bibnamefont {Shikin}}, \ and\ \bibinfo {author} {\bibfnamefont
  {O.}~\bibnamefont {Rader}},\ }\href@noop {} {\bibfield  {journal} {\bibinfo
  {journal} {Nat. Commun.}\ }\textbf {\bibinfo {volume} {3}},\ \bibinfo {pages}
  {1232} (\bibinfo {year} {2012})}\BibitemShut {NoStop}%
\bibitem [{\citenamefont {Bistritzer}\ and\ \citenamefont
  {MacDonald}(2011)}]{Moiré-bands-in-twisted-double-layer-graphene}%
  \BibitemOpen
  \bibfield  {author} {\bibinfo {author} {\bibfnamefont {R.}~\bibnamefont
  {Bistritzer}}\ and\ \bibinfo {author} {\bibfnamefont {A.~H.}\ \bibnamefont
  {MacDonald}},\ }\href@noop {} {\bibfield  {journal} {\bibinfo  {journal}
  {Proceedings of the National Academy of Sciences}\ }\textbf {\bibinfo
  {volume} {108}},\ \bibinfo {pages} {12233} (\bibinfo {year}
  {2011})}\BibitemShut {NoStop}%
\bibitem [{\citenamefont {Xue}\ \emph {et~al.}(2011)\citenamefont {Xue},
  \citenamefont {Sanchez-Yamagishi}, \citenamefont {Bulmash}, \citenamefont
  {Jacquod}, \citenamefont {Deshpande}, \citenamefont {Watanabe}, \citenamefont
  {Taniguchi}, \citenamefont {Jarillo-Herrero},\ and\ \citenamefont
  {LeRoy}}]{JarrilloHerrero}%
  \BibitemOpen
  \bibfield  {author} {\bibinfo {author} {\bibfnamefont {J.}~\bibnamefont
  {Xue}}, \bibinfo {author} {\bibfnamefont {J.}~\bibnamefont
  {Sanchez-Yamagishi}}, \bibinfo {author} {\bibfnamefont {D.}~\bibnamefont
  {Bulmash}}, \bibinfo {author} {\bibfnamefont {P.}~\bibnamefont {Jacquod}},
  \bibinfo {author} {\bibfnamefont {A.}~\bibnamefont {Deshpande}}, \bibinfo
  {author} {\bibfnamefont {K.}~\bibnamefont {Watanabe}}, \bibinfo {author}
  {\bibfnamefont {T.}~\bibnamefont {Taniguchi}}, \bibinfo {author}
  {\bibfnamefont {P.}~\bibnamefont {Jarillo-Herrero}}, \ and\ \bibinfo {author}
  {\bibfnamefont {B.~J.}\ \bibnamefont {LeRoy}},\ }\href {\doibase
  10.1038/nmat2968} {\bibfield  {journal} {\bibinfo  {journal} {Nature
  Materials}\ }\textbf {\bibinfo {volume} {10}},\ \bibinfo {pages} {282}
  (\bibinfo {year} {2011})}\BibitemShut {NoStop}%
\bibitem [{\citenamefont {Moon}\ and\ \citenamefont {Koshino}(2014)}]{mori2}%
  \BibitemOpen
  \bibfield  {author} {\bibinfo {author} {\bibfnamefont {P.}~\bibnamefont
  {Moon}}\ and\ \bibinfo {author} {\bibfnamefont {M.}~\bibnamefont {Koshino}},\
  }\href {\doibase 10.1103/PhysRevB.90.155406} {\bibfield  {journal} {\bibinfo
  {journal} {Phys. Rev. B}\ }\textbf {\bibinfo {volume} {90}},\ \bibinfo
  {pages} {155406} (\bibinfo {year} {2014})}\BibitemShut {NoStop}%
\bibitem [{\citenamefont {McCann}\ and\ \citenamefont
  {Koshino}(2013)}]{bilayer}%
  \BibitemOpen
  \bibfield  {author} {\bibinfo {author} {\bibfnamefont {E.}~\bibnamefont
  {McCann}}\ and\ \bibinfo {author} {\bibfnamefont {M.}~\bibnamefont
  {Koshino}},\ }\href {http://stacks.iop.org/0034-4885/76/i=5/a=056503}
  {\bibfield  {journal} {\bibinfo  {journal} {Reports on Progress in Physics}\
  }\textbf {\bibinfo {volume} {76}},\ \bibinfo {pages} {056503} (\bibinfo
  {year} {2013})}\BibitemShut {NoStop}%
\bibitem [{\citenamefont {Ren}\ \emph {et~al.}(2015)\citenamefont {Ren},
  \citenamefont {Deng}, \citenamefont {Qiao}, \citenamefont {Li}, \citenamefont
  {Jung}, \citenamefont {Zeng}, \citenamefont {Zhang},\ and\ \citenamefont
  {Niu}}]{valley}%
  \BibitemOpen
  \bibfield  {author} {\bibinfo {author} {\bibfnamefont {Y.}~\bibnamefont
  {Ren}}, \bibinfo {author} {\bibfnamefont {X.}~\bibnamefont {Deng}}, \bibinfo
  {author} {\bibfnamefont {Z.}~\bibnamefont {Qiao}}, \bibinfo {author}
  {\bibfnamefont {C.}~\bibnamefont {Li}}, \bibinfo {author} {\bibfnamefont
  {J.}~\bibnamefont {Jung}}, \bibinfo {author} {\bibfnamefont {C.}~\bibnamefont
  {Zeng}}, \bibinfo {author} {\bibfnamefont {Z.}~\bibnamefont {Zhang}}, \ and\
  \bibinfo {author} {\bibfnamefont {Q.}~\bibnamefont {Niu}},\ }\href {\doibase
  10.1103/PhysRevB.91.245415} {\bibfield  {journal} {\bibinfo  {journal} {Phys.
  Rev. B}\ }\textbf {\bibinfo {volume} {91}},\ \bibinfo {pages} {245415}
  (\bibinfo {year} {2015})}\BibitemShut {NoStop}%
\bibitem [{\citenamefont {Gorbachev}\ \emph {et~al.}(2014)\citenamefont
  {Gorbachev}, \citenamefont {Song}, \citenamefont {Yu}, \citenamefont
  {Kretinin}, \citenamefont {Withers}, \citenamefont {Cao}, \citenamefont
  {Mishchenko}, \citenamefont {Grigorieva}, \citenamefont {Novoselov},
  \citenamefont {Levitov},\ and\ \citenamefont {Geim}}]{Gorbachev448}%
  \BibitemOpen
  \bibfield  {author} {\bibinfo {author} {\bibfnamefont {R.}~\bibnamefont
  {Gorbachev}}, \bibinfo {author} {\bibfnamefont {J.}~\bibnamefont {Song}},
  \bibinfo {author} {\bibfnamefont {G.}~\bibnamefont {Yu}}, \bibinfo {author}
  {\bibfnamefont {A.}~\bibnamefont {Kretinin}}, \bibinfo {author}
  {\bibfnamefont {F.}~\bibnamefont {Withers}}, \bibinfo {author} {\bibfnamefont
  {Y.}~\bibnamefont {Cao}}, \bibinfo {author} {\bibfnamefont {A.}~\bibnamefont
  {Mishchenko}}, \bibinfo {author} {\bibfnamefont {I.}~\bibnamefont
  {Grigorieva}}, \bibinfo {author} {\bibfnamefont {K.}~\bibnamefont
  {Novoselov}}, \bibinfo {author} {\bibfnamefont {L.}~\bibnamefont {Levitov}},
  \ and\ \bibinfo {author} {\bibfnamefont {A.~K.}\ \bibnamefont {Geim}},\
  }\href@noop {} {\bibfield  {journal} {\bibinfo  {journal} {Science}\ }\textbf
  {\bibinfo {volume} {346}},\ \bibinfo {pages} {448} (\bibinfo {year}
  {2014})}\BibitemShut {NoStop}%
\bibitem [{\citenamefont {Hermann}(2012)}]{mori1}%
  \BibitemOpen
  \bibfield  {author} {\bibinfo {author} {\bibfnamefont {K.}~\bibnamefont
  {Hermann}},\ }\href {http://stacks.iop.org/0953-8984/24/i=31/a=314210}
  {\bibfield  {journal} {\bibinfo  {journal} {J. Phys.: Cond. Mat.}\ }\textbf
  {\bibinfo {volume} {24}},\ \bibinfo {pages} {314210} (\bibinfo {year}
  {2012})}\BibitemShut {NoStop}%
\bibitem [{\citenamefont {Zhang}\ \emph {et~al.}(2014)\citenamefont {Zhang},
  \citenamefont {Triola},\ and\ \citenamefont {Rossi}}]{mori3}%
  \BibitemOpen
  \bibfield  {author} {\bibinfo {author} {\bibfnamefont {J.}~\bibnamefont
  {Zhang}}, \bibinfo {author} {\bibfnamefont {C.}~\bibnamefont {Triola}}, \
  and\ \bibinfo {author} {\bibfnamefont {E.}~\bibnamefont {Rossi}},\ }\href
  {\doibase 10.1103/PhysRevLett.112.096802} {\bibfield  {journal} {\bibinfo
  {journal} {Phys. Rev. Lett.}\ }\textbf {\bibinfo {volume} {112}},\ \bibinfo
  {pages} {096802} (\bibinfo {year} {2014})}\BibitemShut {NoStop}%
\bibitem [{\citenamefont {Geim}\ and\ \citenamefont
  {Grigorieva}(2013)}]{vanderWaalsH}%
  \BibitemOpen
  \bibfield  {author} {\bibinfo {author} {\bibfnamefont {A.~K.}\ \bibnamefont
  {Geim}}\ and\ \bibinfo {author} {\bibfnamefont {I.~V.}\ \bibnamefont
  {Grigorieva}},\ }\href {http://dx.doi.org/10.1038/nature12385} {\bibfield
  {journal} {\bibinfo  {journal} {Nature}\ }\textbf {\bibinfo {volume} {499}},\
  \bibinfo {pages} {419} (\bibinfo {year} {2013})}\BibitemShut {NoStop}%
\bibitem [{\citenamefont {Jiang}(2015)}]{mog}%
  \BibitemOpen
  \bibfield  {author} {\bibinfo {author} {\bibfnamefont {J.-W.}\ \bibnamefont
  {Jiang}},\ }\href@noop {} {\bibfield  {journal} {\bibinfo  {journal}
  {Frontiers of Physics}\ }\textbf {\bibinfo {volume} {10}},\ \bibinfo {pages}
  {287} (\bibinfo {year} {2015})}\BibitemShut {NoStop}%
\bibitem [{\citenamefont {Gmitra}\ \emph {et~al.}(2016)\citenamefont {Gmitra},
  \citenamefont {Kochan}, \citenamefont {H\"ogl},\ and\ \citenamefont
  {Fabian}}]{mogrf}%
  \BibitemOpen
  \bibfield  {author} {\bibinfo {author} {\bibfnamefont {M.}~\bibnamefont
  {Gmitra}}, \bibinfo {author} {\bibfnamefont {D.}~\bibnamefont {Kochan}},
  \bibinfo {author} {\bibfnamefont {P.}~\bibnamefont {H\"ogl}}, \ and\ \bibinfo
  {author} {\bibfnamefont {J.}~\bibnamefont {Fabian}},\ }\href {\doibase
  10.1103/PhysRevB.93.155104} {\bibfield  {journal} {\bibinfo  {journal} {Phys.
  Rev. B}\ }\textbf {\bibinfo {volume} {93}},\ \bibinfo {pages} {155104}
  (\bibinfo {year} {2016})}\BibitemShut {NoStop}%
\bibitem [{\citenamefont {Wang}\ \emph {et~al.}(2015)\citenamefont {Wang},
  \citenamefont {Ki}, \citenamefont {Chen}, \citenamefont {Berger},
  \citenamefont {MacDonald},\ and\ \citenamefont {Morpurgo}}]{w2}%
  \BibitemOpen
  \bibfield  {author} {\bibinfo {author} {\bibfnamefont {Z.}~\bibnamefont
  {Wang}}, \bibinfo {author} {\bibfnamefont {D.-K.}\ \bibnamefont {Ki}},
  \bibinfo {author} {\bibfnamefont {H.}~\bibnamefont {Chen}}, \bibinfo {author}
  {\bibfnamefont {H.}~\bibnamefont {Berger}}, \bibinfo {author} {\bibfnamefont
  {A.~H.}\ \bibnamefont {MacDonald}}, \ and\ \bibinfo {author} {\bibfnamefont
  {A.~F.}\ \bibnamefont {Morpurgo}},\ }\href@noop {} {\bibfield  {journal}
  {\bibinfo  {journal} {Nat. commun.}\ }\textbf {\bibinfo {volume} {6}}
  (\bibinfo {year} {2015})}\BibitemShut {NoStop}%
\bibitem [{\citenamefont {Gmitra}\ and\ \citenamefont {Fabian}(2015)}]{mogrf2}%
  \BibitemOpen
  \bibfield  {author} {\bibinfo {author} {\bibfnamefont {M.}~\bibnamefont
  {Gmitra}}\ and\ \bibinfo {author} {\bibfnamefont {J.}~\bibnamefont
  {Fabian}},\ }\href {\doibase 10.1103/PhysRevB.92.155403} {\bibfield
  {journal} {\bibinfo  {journal} {Phys. Rev. B}\ }\textbf {\bibinfo {volume}
  {92}},\ \bibinfo {pages} {155403} (\bibinfo {year} {2015})}\BibitemShut
  {NoStop}%
\bibitem [{\citenamefont {Xiao}\ \emph {et~al.}(2012)\citenamefont {Xiao},
  \citenamefont {Liu}, \citenamefont {Feng}, \citenamefont {Xu},\ and\
  \citenamefont {Yao}}]{mos2}%
  \BibitemOpen
  \bibfield  {author} {\bibinfo {author} {\bibfnamefont {D.}~\bibnamefont
  {Xiao}}, \bibinfo {author} {\bibfnamefont {G.-B.}\ \bibnamefont {Liu}},
  \bibinfo {author} {\bibfnamefont {W.}~\bibnamefont {Feng}}, \bibinfo {author}
  {\bibfnamefont {X.}~\bibnamefont {Xu}}, \ and\ \bibinfo {author}
  {\bibfnamefont {W.}~\bibnamefont {Yao}},\ }\href {\doibase
  10.1103/PhysRevLett.108.196802} {\bibfield  {journal} {\bibinfo  {journal}
  {Phys. Rev. Lett.}\ }\textbf {\bibinfo {volume} {108}},\ \bibinfo {pages}
  {196802} (\bibinfo {year} {2012})}\BibitemShut {NoStop}%
\bibitem [{\citenamefont {Liu}\ \emph {et~al.}(2013)\citenamefont {Liu},
  \citenamefont {Shan}, \citenamefont {Yao}, \citenamefont {Yao},\ and\
  \citenamefont {Xiao}}]{mos23}%
  \BibitemOpen
  \bibfield  {author} {\bibinfo {author} {\bibfnamefont {G.-B.}\ \bibnamefont
  {Liu}}, \bibinfo {author} {\bibfnamefont {W.-Y.}\ \bibnamefont {Shan}},
  \bibinfo {author} {\bibfnamefont {Y.}~\bibnamefont {Yao}}, \bibinfo {author}
  {\bibfnamefont {W.}~\bibnamefont {Yao}}, \ and\ \bibinfo {author}
  {\bibfnamefont {D.}~\bibnamefont {Xiao}},\ }\href {\doibase
  10.1103/PhysRevB.88.085433} {\bibfield  {journal} {\bibinfo  {journal} {Phys.
  Rev. B}\ }\textbf {\bibinfo {volume} {88}},\ \bibinfo {pages} {085433}
  (\bibinfo {year} {2013})}\BibitemShut {NoStop}%
\bibitem [{\citenamefont {Rose}\ \emph {et~al.}(2013)\citenamefont {Rose},
  \citenamefont {Goerbig},\ and\ \citenamefont {Pi\'echon}}]{mos24}%
  \BibitemOpen
  \bibfield  {author} {\bibinfo {author} {\bibfnamefont {F.}~\bibnamefont
  {Rose}}, \bibinfo {author} {\bibfnamefont {M.~O.}\ \bibnamefont {Goerbig}}, \
  and\ \bibinfo {author} {\bibfnamefont {F.}~\bibnamefont {Pi\'echon}},\ }\href
  {\doibase 10.1103/PhysRevB.88.125438} {\bibfield  {journal} {\bibinfo
  {journal} {Phys. Rev. B}\ }\textbf {\bibinfo {volume} {88}},\ \bibinfo
  {pages} {125438} (\bibinfo {year} {2013})}\BibitemShut {NoStop}%
\bibitem [{\citenamefont {Lu}\ \emph {et~al.}(2014)\citenamefont {Lu},
  \citenamefont {Li}, \citenamefont {Watanabe}, \citenamefont {Taniguchi},\
  and\ \citenamefont {Andrei}}]{G-MOS2-exp}%
  \BibitemOpen
  \bibfield  {author} {\bibinfo {author} {\bibfnamefont {C.-P.}\ \bibnamefont
  {Lu}}, \bibinfo {author} {\bibfnamefont {G.}~\bibnamefont {Li}}, \bibinfo
  {author} {\bibfnamefont {K.}~\bibnamefont {Watanabe}}, \bibinfo {author}
  {\bibfnamefont {T.}~\bibnamefont {Taniguchi}}, \ and\ \bibinfo {author}
  {\bibfnamefont {E.~Y.}\ \bibnamefont {Andrei}},\ }\href {\doibase
  10.1103/PhysRevLett.113.156804} {\bibfield  {journal} {\bibinfo  {journal}
  {Phys. Rev. Lett.}\ }\textbf {\bibinfo {volume} {113}},\ \bibinfo {pages}
  {156804} (\bibinfo {year} {2014})}\BibitemShut {NoStop}%
\bibitem [{\citenamefont {Avsar}\ \emph {et~al.}(2014)\citenamefont {Avsar},
  \citenamefont {Tan}, \citenamefont {Taychatanapat}, \citenamefont
  {Balakrishnan}, \citenamefont {Koon}, \citenamefont {Yeo}, \citenamefont
  {Lahiri}, \citenamefont {Carvalho}, \citenamefont {Rodin}, \citenamefont
  {O’Farrell} \emph {et~al.}}]{g-ws2}%
  \BibitemOpen
  \bibfield  {author} {\bibinfo {author} {\bibfnamefont {A.}~\bibnamefont
  {Avsar}}, \bibinfo {author} {\bibfnamefont {J.~Y.}\ \bibnamefont {Tan}},
  \bibinfo {author} {\bibfnamefont {T.}~\bibnamefont {Taychatanapat}}, \bibinfo
  {author} {\bibfnamefont {J.}~\bibnamefont {Balakrishnan}}, \bibinfo {author}
  {\bibfnamefont {G.}~\bibnamefont {Koon}}, \bibinfo {author} {\bibfnamefont
  {Y.}~\bibnamefont {Yeo}}, \bibinfo {author} {\bibfnamefont {J.}~\bibnamefont
  {Lahiri}}, \bibinfo {author} {\bibfnamefont {A.}~\bibnamefont {Carvalho}},
  \bibinfo {author} {\bibfnamefont {A.}~\bibnamefont {Rodin}}, \bibinfo
  {author} {\bibfnamefont {E.}~\bibnamefont {O’Farrell}},  \emph {et~al.},\
  }\href@noop {} {\bibfield  {journal} {\bibinfo  {journal} {Nat. commun.}\
  }\textbf {\bibinfo {volume} {5}} (\bibinfo {year} {2014})}\BibitemShut
  {NoStop}%
\bibitem [{\citenamefont {Asmar}\ and\ \citenamefont
  {Ulloa}(2014)}]{mahmoud-PRL}%
  \BibitemOpen
  \bibfield  {author} {\bibinfo {author} {\bibfnamefont {M.~M.}\ \bibnamefont
  {Asmar}}\ and\ \bibinfo {author} {\bibfnamefont {S.~E.}\ \bibnamefont
  {Ulloa}},\ }\href {\doibase 10.1103/PhysRevLett.112.136602} {\bibfield
  {journal} {\bibinfo  {journal} {Phys. Rev. Lett.}\ }\textbf {\bibinfo
  {volume} {112}},\ \bibinfo {pages} {136602} (\bibinfo {year}
  {2014})}\BibitemShut {NoStop}%
\bibitem [{\citenamefont {Asmar}\ and\ \citenamefont
  {Ulloa}(2015)}]{mahmoud-PRB}%
  \BibitemOpen
  \bibfield  {author} {\bibinfo {author} {\bibfnamefont {M.~M.}\ \bibnamefont
  {Asmar}}\ and\ \bibinfo {author} {\bibfnamefont {S.~E.}\ \bibnamefont
  {Ulloa}},\ }\href {\doibase 10.1103/PhysRevB.91.165407} {\bibfield  {journal}
  {\bibinfo  {journal} {Phys. Rev. B}\ }\textbf {\bibinfo {volume} {91}},\
  \bibinfo {pages} {165407} (\bibinfo {year} {2015})}\BibitemShut {NoStop}%
\bibitem [{\citenamefont {Xiao}\ \emph {et~al.}(2010)\citenamefont {Xiao},
  \citenamefont {Chang},\ and\ \citenamefont {Niu}}]{berry}%
  \BibitemOpen
  \bibfield  {author} {\bibinfo {author} {\bibfnamefont {D.}~\bibnamefont
  {Xiao}}, \bibinfo {author} {\bibfnamefont {M.-C.}\ \bibnamefont {Chang}}, \
  and\ \bibinfo {author} {\bibfnamefont {Q.}~\bibnamefont {Niu}},\ }\href
  {\doibase 10.1103/RevModPhys.82.1959} {\bibfield  {journal} {\bibinfo
  {journal} {Rev. Mod. Phys.}\ }\textbf {\bibinfo {volume} {82}},\ \bibinfo
  {pages} {1959} (\bibinfo {year} {2010})}\BibitemShut {NoStop}%
\bibitem [{\citenamefont {Qiao}\ \emph {et~al.}(2013)\citenamefont {Qiao},
  \citenamefont {Li}, \citenamefont {Tse}, \citenamefont {Jiang}, \citenamefont
  {Yao},\ and\ \citenamefont {Niu}}]{graphene-Chern}%
  \BibitemOpen
  \bibfield  {author} {\bibinfo {author} {\bibfnamefont {Z.}~\bibnamefont
  {Qiao}}, \bibinfo {author} {\bibfnamefont {X.}~\bibnamefont {Li}}, \bibinfo
  {author} {\bibfnamefont {W.-K.}\ \bibnamefont {Tse}}, \bibinfo {author}
  {\bibfnamefont {H.}~\bibnamefont {Jiang}}, \bibinfo {author} {\bibfnamefont
  {Y.}~\bibnamefont {Yao}}, \ and\ \bibinfo {author} {\bibfnamefont
  {Q.}~\bibnamefont {Niu}},\ }\href@noop {} {\bibfield  {journal} {\bibinfo
  {journal} {Phys. Rev. B}\ }\textbf {\bibinfo {volume} {87}},\ \bibinfo
  {pages} {125405} (\bibinfo {year} {2013})}\BibitemShut {NoStop}%
\bibitem [{\citenamefont {Hao}\ \emph {et~al.}(2008)\citenamefont {Hao},
  \citenamefont {Zhang}, \citenamefont {Wang}, \citenamefont {Zhang},\ and\
  \citenamefont {Wang}}]{ribons2}%
  \BibitemOpen
  \bibfield  {author} {\bibinfo {author} {\bibfnamefont {N.}~\bibnamefont
  {Hao}}, \bibinfo {author} {\bibfnamefont {P.}~\bibnamefont {Zhang}}, \bibinfo
  {author} {\bibfnamefont {Z.}~\bibnamefont {Wang}}, \bibinfo {author}
  {\bibfnamefont {W.}~\bibnamefont {Zhang}}, \ and\ \bibinfo {author}
  {\bibfnamefont {Y.}~\bibnamefont {Wang}},\ }\href {\doibase
  10.1103/PhysRevB.78.075438} {\bibfield  {journal} {\bibinfo  {journal} {Phys.
  Rev. B}\ }\textbf {\bibinfo {volume} {78}},\ \bibinfo {pages} {075438}
  (\bibinfo {year} {2008})}\BibitemShut {NoStop}%
\bibitem [{\citenamefont {Li}\ \emph {et~al.}(2011)\citenamefont {Li},
  \citenamefont {Martin}, \citenamefont {B\"{u}ttiker},\ and\ \citenamefont
  {Morpurgo}}]{Li2011}%
  \BibitemOpen
  \bibfield  {author} {\bibinfo {author} {\bibfnamefont {J.}~\bibnamefont
  {Li}}, \bibinfo {author} {\bibfnamefont {I.}~\bibnamefont {Martin}}, \bibinfo
  {author} {\bibfnamefont {M.}~\bibnamefont {B\"{u}ttiker}}, \ and\ \bibinfo
  {author} {\bibfnamefont {A.~F.}\ \bibnamefont {Morpurgo}},\ }\href@noop {}
  {\bibfield  {journal} {\bibinfo  {journal} {Nat. Phys.}\ }\textbf {\bibinfo
  {volume} {7}},\ \bibinfo {pages} {38} (\bibinfo {year} {2011})}\BibitemShut
  {NoStop}%
\bibitem [{\citenamefont {Segarra}\ \emph {et~al.}(2016)\citenamefont
  {Segarra}, \citenamefont {Planelles},\ and\ \citenamefont
  {Ulloa}}]{CarlosEdges}%
  \BibitemOpen
  \bibfield  {author} {\bibinfo {author} {\bibfnamefont {C.}~\bibnamefont
  {Segarra}}, \bibinfo {author} {\bibfnamefont {J.}~\bibnamefont {Planelles}},
  \ and\ \bibinfo {author} {\bibfnamefont {S.~E.}\ \bibnamefont {Ulloa}},\
  }\href {\doibase 10.1103/PhysRevB.93.085312} {\bibfield  {journal} {\bibinfo
  {journal} {Phys. Rev. B}\ }\textbf {\bibinfo {volume} {93}},\ \bibinfo
  {pages} {085312} (\bibinfo {year} {2016})}\BibitemShut {NoStop}%
\bibitem [{\citenamefont {Qiao}\ \emph {et~al.}(2012)\citenamefont {Qiao},
  \citenamefont {Jiang}, \citenamefont {Li}, \citenamefont {Yao},\ and\
  \citenamefont {Niu}}]{theory}%
  \BibitemOpen
  \bibfield  {author} {\bibinfo {author} {\bibfnamefont {Z.}~\bibnamefont
  {Qiao}}, \bibinfo {author} {\bibfnamefont {H.}~\bibnamefont {Jiang}},
  \bibinfo {author} {\bibfnamefont {X.}~\bibnamefont {Li}}, \bibinfo {author}
  {\bibfnamefont {Y.}~\bibnamefont {Yao}}, \ and\ \bibinfo {author}
  {\bibfnamefont {Q.}~\bibnamefont {Niu}},\ }\href {\doibase
  10.1103/PhysRevB.85.115439} {\bibfield  {journal} {\bibinfo  {journal} {Phys.
  Rev. B}\ }\textbf {\bibinfo {volume} {85}},\ \bibinfo {pages} {115439}
  (\bibinfo {year} {2012})}\BibitemShut {NoStop}%
\bibitem [{Sup()}]{Suppl}%
  \BibitemOpen
  \href@noop {} {\bibinfo  {journal} {For details of tight binding calculations
  and other TMD models, see supplementary material}\ }\BibitemShut {NoStop}%
\bibitem [{\citenamefont {Konschuh}\ \emph {et~al.}(2010)\citenamefont
  {Konschuh}, \citenamefont {Gmitra},\ and\ \citenamefont {Fabian}}]{G-Rash}%
  \BibitemOpen
\bibfield  {journal} {  }\bibfield  {author} {\bibinfo {author} {\bibfnamefont
  {S.}~\bibnamefont {Konschuh}}, \bibinfo {author} {\bibfnamefont
  {M.}~\bibnamefont {Gmitra}}, \ and\ \bibinfo {author} {\bibfnamefont
  {J.}~\bibnamefont {Fabian}},\ }\href {\doibase 10.1103/PhysRevB.82.245412}
  {\bibfield  {journal} {\bibinfo  {journal} {Phys. Rev. B}\ }\textbf {\bibinfo
  {volume} {82}},\ \bibinfo {pages} {245412} (\bibinfo {year}
  {2010})}\BibitemShut {NoStop}%
\bibitem [{\citenamefont {Slater}\ and\ \citenamefont {Koster}(1954)}]{slater}%
  \BibitemOpen
  \bibfield  {author} {\bibinfo {author} {\bibfnamefont {J.~C.}\ \bibnamefont
  {Slater}}\ and\ \bibinfo {author} {\bibfnamefont {G.~F.}\ \bibnamefont
  {Koster}},\ }\href {\doibase 10.1103/PhysRev.94.1498} {\bibfield  {journal}
  {\bibinfo  {journal} {Phys. Rev.}\ }\textbf {\bibinfo {volume} {94}},\
  \bibinfo {pages} {1498} (\bibinfo {year} {1954})}\BibitemShut {NoStop}%
\bibitem [{Note1()}]{Note1}%
  \BibitemOpen
  \bibinfo {note} {We have kept $t_R$ constant in the tight binding
  calculations, although a gate voltage dependence is likely. This would be
  expected to reduce the gate voltage required to close the gap.}\BibitemShut
  {Stop}%
\end{thebibliography}%
\bibliographystyle{apsrev4-1}

\end{document}